# Applying Remote Handling Attributes to the ITER Neutral Beam Cell Monorail Crane

O Crofts[a], P Allan[a], J Raimbach[a], A Tesini[b], C-H Choi[b], C Damiani[c], M Van Uffelen[c]

*a CCFE. Culham Science Centre, Abingdon, OX14 3DB, UK,*
*b ITER Organisation, CS90 046, 13067 St. Paul les Durance Cedex, France,*
*c Fusion for Energy, C/Josep Pla 2, Torres Diagonal Litoral-B3, E-08019 Barcelona – Spain,*

The maintenance requirements for the equipment in the ITER Neutral Beam Cell requires components to be lifted and transported within the cell by remote means. To meet this requirement, the provision of an overhead crane with remote handling capabilities has been initiated. The layout of the cell has driven the design to consist of a monorail crane that travels on a branched monorail track attached to the cell ceiling.

This paper describes the principle design constraints and how the remote handling attributes were applied to the concept design of the monorail crane, concentrating on areas where novel design solutions have been required and on the remote recovery requirements and solutions.

Keywords: Monorail Crane Remote Handling ITER

## 1. Introduction

The Monorail Crane forms part of the ITER Neutral Beam Cell Remote Handling System, for which the Conceptual Design Review has just been completed. The status of the System design by CCFE is the subject of a paper presented at the SOFT 2012 conference; Sykes [1].

The Monorail Crane is the principal transporter for all plant and equipment within the Neutral Beam Cell and is used during installation and maintenance. The Cell contains up to 3 Heating Neutral Beams, a Diagnostic Neutral Beam and 4 upper ports.

The Neutral Beam Cell contains a series of pillars to support the upper floors of the Tokamak Building. These pillars preclude the use of an X-Y bridge crane. An overhead monorail crane is therefore proposed in the concept design, based on the IBERTEF reference design [2] and is described in detail in the ITER concept Design Description Document [3].

A summary of the remote handling attributes applied to the concept design are presented in this paper.

### 1.1 Principal Design Constraints

The safe working load of the crane is 50t.

Virtual Reality simulations of the crane operations show that the highest hook heights are required when the tall Beam Line Components, such as the Calorimeter and Residual Ion Dump, are lifted over the Balcony Plates.

The height of the components and the distance between the Balcony Plates and the Cell ceiling imposes a tight constraint on the maximum height of the crane. It is a maximum of 1400mm when adhering to the minimum clearance of 100mm applied to all remote crane operations.

The crane requires a four rope lift to accommodate small off-centre loads and to allow accurate position control of components during lifting and lowering. This ensures correct engagement with remote alignment and location features such as dowels.

When shielding or containment barriers have been removed during maintenance, personnel access to the Neutral Beam Cell will not be possible. The crane must therefore be operable and recoverable entirely remotely.

The safety case requires the crane to retain its load during a Seismic Level 2 (SL-2) event.

The ITER System Requirements for the concept design of the Neutral Beam Cell Remote Handling Equipment requires that all Remote Handling Equipment be recoverable by credible means and for all components to have a minimum radiation tolerance of 20kGy.

### 1.2 Design Overview

The Monorail Crane is shown in figure 1 transporting the calorimeter. The crane system comprises; the monorail, upon which run two bogies that are mounted to the crane frame. The crane frame supports four hoist assemblies that raise and lower the lifting frame. Each of these assemblies is described in more detail in the following sections.

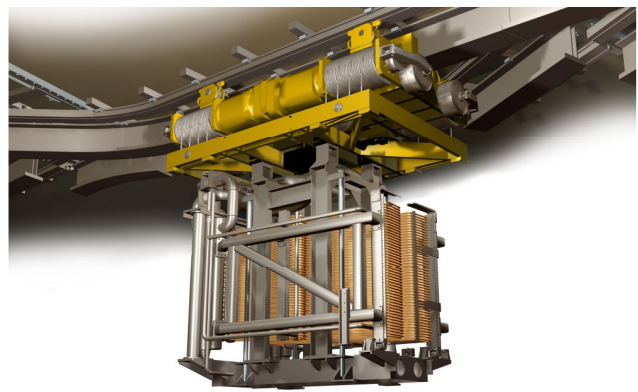

Fig.1 Monorail Crane System.

---

*author's email: Oliver.Crofts@CCFE.ac.uk*

## 2. Monorail

The Neutral Beam Cell monorail is shown in red figure 2. At the top of the figure, the monorail track passes behind the three heating neutral beam lines and at the bottom it passes above the front end components and has branches to pass over each of the three heating beam lines.

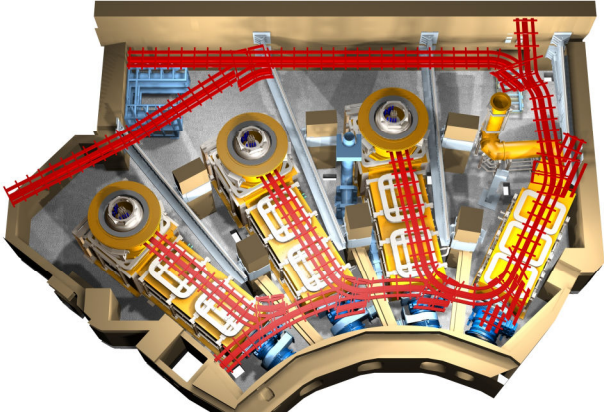

Fig.2 Plan view on the Neutral Beam Cell.

Seven sets of switches allow the crane to move between the different branches of the monorail. The switches run on linear slides driven from the Level 3 High Voltage Deck above.

The monorail design is shown in figure 3 below. It comprises a main central I beam with stabilizer rails to each side to react eccentric loads. These are attached to cross-beams, mounted to plates embedded in the Cell ceiling.

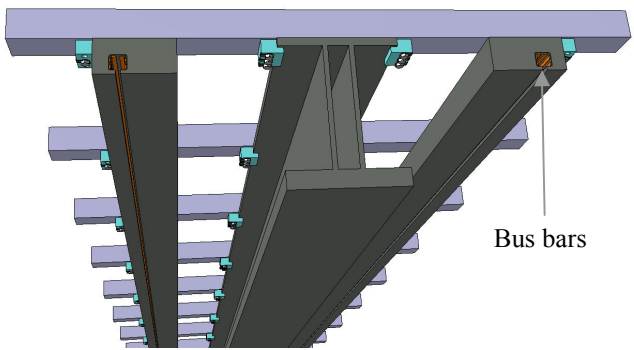

Fig.3 Monorail arrangement.

The two stabilizer rails contain bus bar electrical lines that connect to the crane via multiple pick-ups on the crane bogies to ensure pick-up when crossing switches and to provide redundancy. The bus bars can carry power and signal communication.

## 3. Bogies

Two bogies support the crane on the monorail with a total of four independent drives.

Each bogey has two sprung stabilizer wheels with Ackermann steering to maintain constant contact with the stabilizer rails and four conductor bus pick-up assemblies based on the Demag DCL system to supply electrical power and signals to the crane.

## 4. Hoists

The crane has four independent hoist assemblies, mounted to the crane frame. The assembly comprises; rope drum, drives and brakes, shown in red in figure 4.

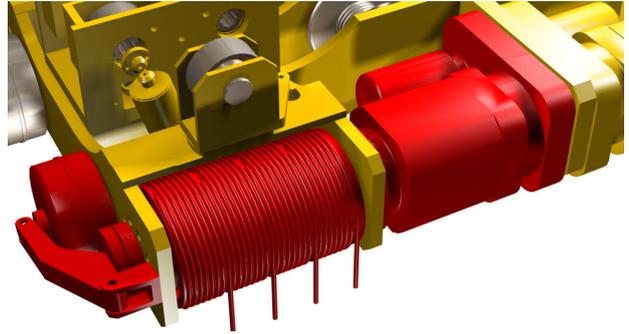

Fig.4 Hoist assembly arrangement.

Due to the restricted vertical height of the crane, the rope drum diameter was limited to 450mm. Single ropes with suitable breaking loads cannot be wound round such a small drum so four rope drops are used on each drum. The rope selected is an 18mm diameter Diepa H50, compacted strand wire rope.

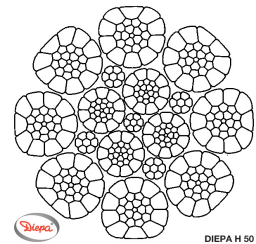

Fig.5 Compacted strand wire rope arrangement.

The hoist drive requires a large speed range to achieve both the operational efficiency requirements and the controlled engagement of components. A Demag 20kW conical rotor motor with integrated 2kW creep motor and duty brake meets these requirements, coupled to a 226:1 three stage planetary gearbox packaged as one assembly inside the rope drum.

The European Standard for crane safety and general design [4] requires an emergency brake that acts directly on the drum. The diameter of a standard disc brake design is too large to fit in the restricted vertical height of the crane so a conical brake has been used at one end of each rope drum, actuated by disc springs and disengaged with a standard crane emergency brake electromagnetic actuator by Stromag.

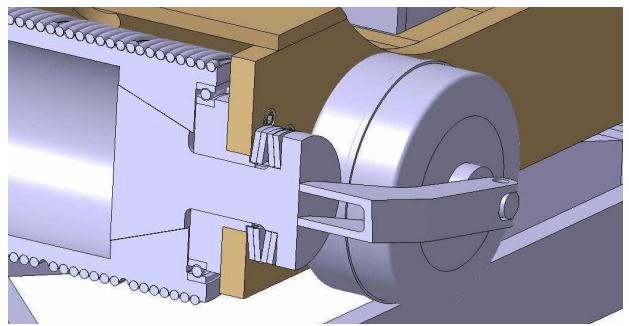

Fig.6 Conical brake and actuator arrangement.

## 5. Lifting Frame

The lifting frame provides the standard lifting interface between the crane and components and it interfaces with lifting adaptors in operations where components require additional motions or a non-standard lifting interface.

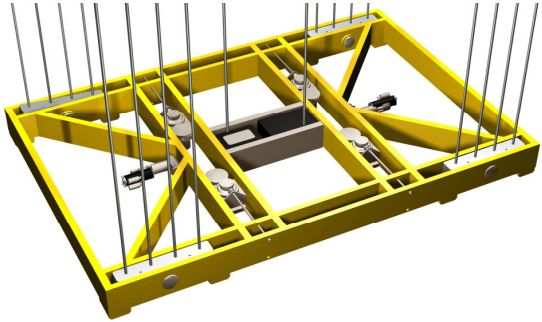

Fig.7 Lifting frame arrangement.

### 5.1 Twist-locks

Mechanical engagement is provided by four twist-locks conforming to international standards [5].

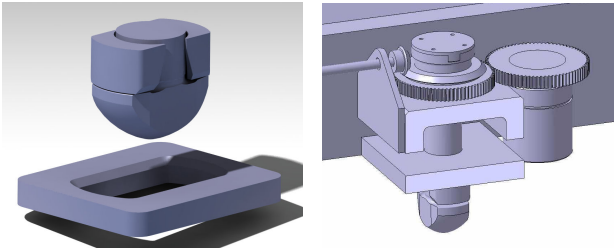

Fig.8 Twist-lock arrangement.

The twist-locks provide alignment during attachment of the lifting frame. They have external drive connections that can be driven by a tool deployed by any of the Cell manipulators in case of motor failure. The entire twist-lock assembly can also be replaced remotely.

### 5.2 Equalizer blocks

The lifting frame is suspended from the crane ropes which pass through equalizer blocks at each corner of the frame.

Within each equalizer block the ropes pass around pulleys on each end of a rocker bar to ensure equal tension in each of the four rope drops, even if the rope creep rate or extension under load varies between drops.

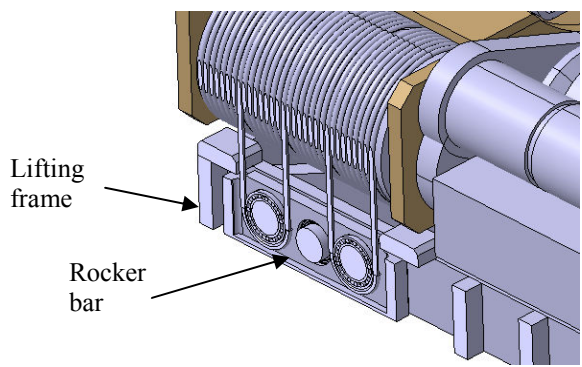

Fig.9 Section through an equalizer block arrangement.

## 6. Control

A unique umbilical control connection to the crane is not possible because the track does not have a single origin and there is no space in the Cell for a reel or festoon. Three other options have been considered for the concept design and these are described below.

### 6.1 CAN bus

This option uses additional bars in the Demag DCL conductor bar power transmission system described above to transmit CAN bus communication signals.

The system is commonly used on production lines but is susceptible to noise and it has a relatively low band width, preventing the use of video cameras on the crane or lifting frame.

The CAN bus system requires onboard processing. Radiation tolerance of the processors is a potential issue. Commercial components are available with radiation tolerance levels up to a few kGy but they are expensive.

The requirement specification states a minimum tolerance of 20kGy. The actual dose received by the crane is likely to be much lower than this but some shielding may be required.

### 6.2 Wireless transmission

This option uses radio signals to send and receive control communication. It has similar issues to the CAN bus system in requiring onboard electronics and has a susceptibility to noise.

### 6.3 Discrete plug-in points

This option uses the DCL power bus connections to directly drive the crane to discrete points along the monorail where it can remotely connect to control plug-in points adjacent to the track.

Flexibility in the connection between the crane and the plug-in point could allow the crane to move a metre or so in either direction along the monorail whilst plugged in. However, a large number of plug-in points would be required and some flexibility of the design would be lost.

### 6.4 Selected option

All options are viable at the concept stage. The discrete plug-in points option has lower development risks than the other options and does not require the same level of radiation tolerant electronics. However, this option has been assessed as considerably more expensive due to the extensive cable and signal management requirements. Therefore, this option should only be considered if neither of the other options can be developed into an acceptable solution.

The CAN bus and wireless transmission options have complementary strengths and weaknesses. The wireless transmission system is being considered for use with the ITER Cask Transfer System and would therefore have reduced development costs and risk and there would be commonality between the ITER control systems.

## 7. Recovery

To achieve the required availability of the ITER Neutral Beam Systems; high reliability components, redundancy, condition monitoring and regular maintenance will be required to ensure the crane is suitably reliable.

In the event of failure when shielding or containment barriers have been removed, remote recovery must be possible. This is achieved with a number of systems, including:

1. The ability to lift a load on two out of the four hoists in the event that one hoist seizes.

2. Torque limiters on the monorail drives to allow the crane to return to the transfer area with one drive seized.

3. Dexterous manipulation available at a number of locations in cell to allow recovery, release or repair of failed components.

4. A recovery hoist system to lower a section of monorail and the crane onto a stillage for removal, in a cask, to the Hot Cell for maintenance.

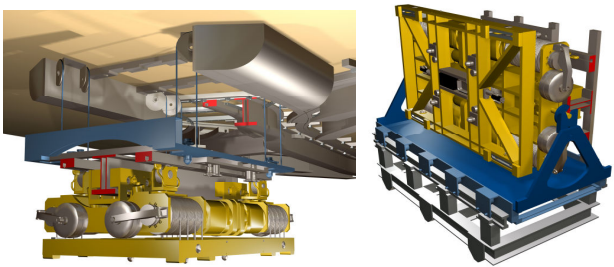

Fig.10 The crane at the recovery hoist position and at the transfer table, rotated through 90 degrees on the stillage to fit into a transfer cask.

The recovery hoist system will provide the preferred method of access to the Crane for planned and unplanned maintenance, whether or not personnel access is possible.

## 8. Seismic Loads

The crane is required not to drop its load during a Seismic Level 2 (SL-2) event. The crane is also required to provide a credible recovery scenario for other Remote Handling Equipment in the Cell following such an event. To this end, the crane has been designed to withstand the event without unrecoverable damage.

The variable natural frequency of the load suspended from the crane due to the varying length of rope during a lift means that for most heavy lifts, there is a point where the natural frequency will match that of the building response to a seismic event. Under these circumstances, during an SL-2 event, the acceleration of the mass would exceed gravity.

When the upward acceleration of the load on the rope exceeds gravity a non-linear slack rope condition arises, where higher rope tensions are seen when the rope becomes taught again, compared to the loads that would be seen if the rope acted as a spring.

Transient dynamic analysis was performed using an iterative small time step calculation on a one dimensional system to show the maximum rope loads for a range of rope lengths and seismic input frequency. The effects of varying rope stiffness and damping was also investigated.

It was found that the maximum rope load for the non-linear system was about 1/3 higher than that for a linear system where the ropes acted as springs.

Structural analysis showed some strengthening of the crane and lifting frame was required to withstand the additional load and that the loads on the building interface were high.

Additional work was carried out to strengthen the crane and to add flexible mounts between the cross-beams and the building interface points to spread the crane load over more building interface points.

Further analysis will be required using more comprehensive input movement data and a multi-degree of freedom model to consider also the effects of a rotating and off-centre load.

## 9. Conclusions

A feasible concept design with all the required Remote Handling attributes has been achieved that meets the system requirements.

Considerable work remains for the preliminary design stage due to the novel nature of some areas of the design, most notably the hoist and control system and also in demonstrating that the requirements of the safety case have been met.

### Acknowledgments

This work was funded jointly by the RCUK Energy Programme under grant EP/I501045 and by Fusion for Energy under grant 2009-GRT-051. The views and opinions expressed herein do not necessarily reflect those of Fusion For Energy or European Commission or ITER Organization. Fusion For Energy is not liable for the use which might be made of the information in this publication